\documentclass[onecolumn,showpacs,showkeys,preprintnumbers,amsmath,amssymb]{revtex4}

\usepackage{graphicx}
\usepackage{dcolumn}
\usepackage{bm}
\usepackage[utf8x]{inputenc}
\usepackage{float}

\begin{document}


\title{Neural Network Study of Hidden-Charm Pentaquark Resonances} 

\author{Halil Mutuk}
\email{halilmutuk@gmail.com}
\affiliation{%
Physics Department, Faculty of Arts and Sciences, Ondokuz Mayis University, 55139, Samsun, Turkey
}%


\begin{abstract}
Very recently, LHCb experiment announced the observation of hidden-charm pentaquark states $P_c(4312)$, $P_c(4440)$, and $P_c(4457)$ near the $ \Sigma_c \bar{D}$ and $ \Sigma_c \bar{D}^\ast$ thresholds, respectively. In this present work, we studied thesepentaquarks in the framework of the nonrelativistic quark model with four types of potential. We solved 5-body Schrödinger equation by using artificial neural network method and made predictions of parities for these states which are not determined in the experiment yet. The mass of another possible pentaquark state near the $\bar{D}^\ast \Sigma_c^\ast$ with $J^P=5/2^-$ is also calculated.
\end{abstract}

\maketitle

\section{\label{sec:level1}Introduction}
In recent years, some experimental states or resonances are announced to be observed to be candidates beyond the conventional quark-antiquark and three-quark configurations. Most of these particles are not confirmed with high statistics and better resolution. Besides that, except the case for $X(3872)$ \cite{1} , they were seen only in one experiment, such as $X(5568)$ \citep{2,3} or in one type of experiment such as $B$ factories. The observation of $X(3872)$ was a milestone for the era of so called exotic states. Exotic states are beyond the description of conventional quark model. Pentaquark is an example of these exotic states. It consists of four quarks ($qqqq$) and one antiquark ($\bar{q}$) bound together. 

The situation turned into a new perspective with the first discovery of the pentaquark candidates, $P_c(4450)$ and $P_c(4380)$ by LHCb in 2015 \cite{4}. There were theoretical studies for these pentaquark particles prior to their observation \cite{5,6,7,8}. The masses of these states were very close to $\bar{D}^* \Sigma_c^*$ threshold. This makes comfortable to assume that those two pentaquarks as baryon-meson molecule \cite{9,10,11,12,13,14,15,16,17,18,19}. The other possibilities are compact pentaquark \cite{20,21,22,23}, quark model \cite{24,25,26}, chiral quark model \cite{27}, quark-cluster model \cite{28} and baryocharmonium model \cite{29}. 

Very recently, the LHCb collaboration updated the results of Ref. \cite{4} reporting the observation of new narrow pentaquark states \cite{30} with masses and widths as follows:

\begin{eqnarray*}
P_c(4312) ~ M &=& (4311.9 \pm 0.7^{+6.8}_{-0.6}) ~ \text{MeV},  \\
\Gamma &=& (9.8 \pm 2.7^{+3.7}_{-4.5}) ~ \text{MeV}, \\
P_c(4440) ~ M &=& (4440.3 \pm 1.3^{+4.1}_{-4.7}) ~ \text{MeV},  \\
\Gamma &=& (20.6 \pm 4.9^{+8.7}_{-10.1}) ~ \text{MeV}, \\
P_c(4457) ~ M &=& (4457.3 \pm 0.6^{+4.1}_{-1.7}) ~ \text{MeV},  \\
\Gamma &=& (6.4 \pm 2.0^{+5.7}_{-1.9}) ~ \text{MeV}.
\end{eqnarray*}

 The massess of  $P_c(4440)$  and  $P_c(4457)$ are close to $\Sigma_c \bar{D}^*$  threshold and the mass of $P_c(4312)$ is very close to $ \Sigma_c \bar{D}$ threshold. As pointed out in \cite{31}, central mass of the $P_c(4312)$ state is $\approx$6 MeV below the $\Sigma_c^+\bar{D}^0$ threshold and $\approx$12 MeV below the $\Sigma_c^{++}D^-$ threshold. For $P_c(4440)$, it is $\approx$20 MeV below the $\Sigma_c^+\bar{D}^{*0}$ and $\approx$24 MeV below the $\Sigma_c^{++}\bar{D}^{*-}$ thresholds. In the case of $P_c(4457)$, it is 
$\approx$3 MeV below the $\Sigma_c^+\bar{D}^{*0}$ and $\approx$7 MeV below the $\Sigma_c^{++}\bar{D}^{*-}$ thresholds. Isospin violating process can occur when the width of a resonance is small and mass is below the corresponding thresholds. This can be instance for these pentaquarks. 

The observation of these pentaquarks got attention immediately \cite{32,33,34,35,36,37,38,39}. In this paper, we use constituent quark model in order to obtain spectrum and quantum numbers. As mentioned in Ref. \cite{25}, constituent quark model has often been employed  for exploratory studies in QCD and paved the way for lattice simulations and QCD sum rules calculations. The main part of the constituent quark model is to get a solution of Schrödinger equation with a specific potential. For mesons and baryons, this can be done effectively and one can obtain reliable results comparing to the results of experiments. But pentaquark structures are multiquark systems and due to the complex interactions among quarks,  solving 5-body Schrödinger equation is a challenging task. For this purpose, we solved Schrödinger equation via Artificial Neural Network (ANN). 

Besides other use of fields, ANNs can be utilized as an elective strategy to solve differential conditions and quantum mechanical systems \cite{40,41}. ANNs provide some advantages compared to standard numerical methods \cite{42,43}
\begin{itemize}
   \item The solution is continuous over all the domain of integration,
   \item With the number of sampling points and dimensions of the problem, computational complexity does not increase significantly,
   \item Rounding-off error propagation of standard numerical methods does not influence the neural network solution,
   \item The method requires less number of model parameters and therefore does not ask for high memory space in computer.
\end{itemize}

The paper is organized as follows. In Section \ref{sec:level2}, the model and method used for the calculations are described. In Section \ref{sec:level3}, obtained results are discussed and in Section \ref{sec:level4}, we sum up our work.

\section{\label{sec:level2}The Model and Method}
\subsection{The Model} 
The Hamiltonian of Ref. \cite{44} reads as follows
\begin{equation}
H=\sum_i \left( m_i+\frac{\bf{p}_i^2}{2m_i} \right)-\frac{3}{16}\sum_{i<j} \tilde{\lambda}_i \tilde{\lambda}_j v_{ij}(r_{ij})
\end{equation}
with the potential 
\begin{eqnarray}
v_{ij}(r)&=&-\frac{\kappa(1-e^{-\frac{r}{r_c}})}{r}+\lambda r^p + \Lambda +
           \frac{2\pi}{3m_im_j}\kappa^\prime (1-e^{-\frac{r}{r_c}})\frac{e^{-\frac{r^2}{r_0^2}}}{\pi^{3/2}r_0^3} \bf{\sigma_i} \bf{\sigma_j}, \label{potential}
\end{eqnarray}
where $r_0(m_i,m_j)=A \left(\frac{2m_im_j}{m_i+m_j} \right)^{-B}$, $A$ and $B$ are constant parameters, $\kappa$ and $\kappa^\prime$ are parameters, $r_{ij}$ is the interquark distance $\vert \bf{r_i}-\bf{r_j \vert}$, $\sigma_i$ are the Pauli matrices and $\tilde{\lambda}_i$ are Gell-Mann matrices. There are four potentials referred to the $p$ nd $r_c$:
\begin{eqnarray*}
\text{AL1} &\to & p=1, ~ r_c=0, \\
\text{AP1} &\to & p=2/3, ~ r_c=0, \\
\text{AL2} &\to & p=1, ~ r_c \neq 0, \\
\text{AP2} &\to & p=2/3,  r_c \neq 0.
\end{eqnarray*}
The related parameters are given in Table \ref{tab:table1}.

\begin{table}[H]
\caption{\label{tab:table1}Parameters of the potentials.}
\begin{ruledtabular}
\begin{tabular}{cccccc}
 &AL1&AP1&AL2& AP2\\
\hline
$m_u=m_d$& 0.315 GeV & 0.277 GeV & 0.320 GeV &0.280 GeV \\
 $m_s$& 0.577 GeV & 0.553 GeV & 0.587 GeV &0.569 GeV \\
$m_c$ & 1.836 GeV & 1.819 GeV  & 1.851 GeV &1.840 GeV \\
$m_b$ & 5.227 GeV & 5.206 GeV & 5.231 GeV &5.213 GeV \\
$\kappa$ & 0.5069 & 0.4242 & 0.5871 &0.5743 \\
$\kappa^\prime$ & 1.8609 & 1.8025  & 1.8475  &1.8993 \\
$\lambda$& 0.1653 $\text{GeV}^2$ & 0.3898 $\text{GeV}^{5/3}$ & 0.1673 $\text{GeV}^2$ &0.3978 $\text{GeV}^{5/3}$  \\
$\Lambda$& -0.8321 GeV & -1.1313 GeV & -0.8182 GeV & -1.1146 GeV \\
$B$ &0.2204 &0.3263& 0.2132&0.3478\\
$A$& 1.6553 $\text{GeV}^{B-1}$ & 1.5296 $\text{GeV}^{B-1}$& 1.6560 $\text{GeV}^{B-1}$ & 1.5321 $\text{GeV}^{B-1}$\\
$r_c$ &0 &0& 0.1844 $\text{GeV}^{-1}$ & 0.3466 $\text{GeV}^{-1}$
\end{tabular}
\end{ruledtabular}
\end{table}
This potential was developed under the nonrelativistic quark model (NRQM) and used for exploratory studies. It compose of 'Coulomb + linear' or 'Coulomb + 2/3-power'  term and a strong but smooth hyperfine term. For further details of this potential, see Ref. \cite{44}. They built a new interquark potential which work on meson and baryon sector equally well. This simple quark model is based on nonrelativistic kinetic energy and a color-additive interaction related to pairwise forces carried by color-octet exchanges \cite{25}.

\subsection{The Method}
\begin{figure}[H]
\centering
\includegraphics[width=3.4in]{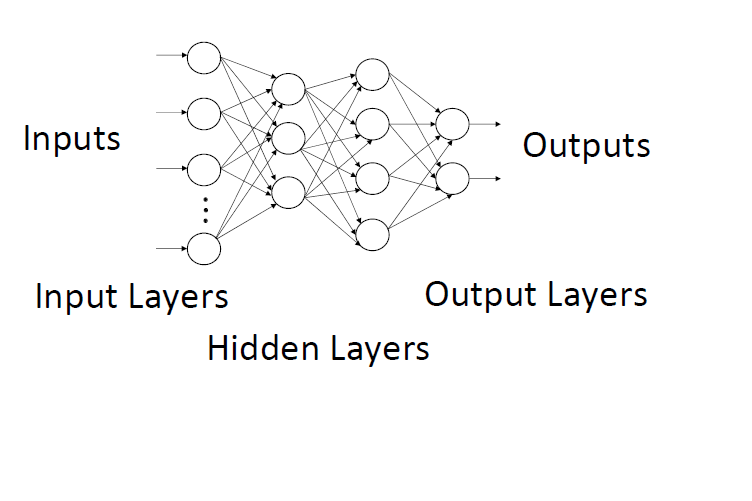}
\caption{\label{fig:1}A model of multilayer neural networks}
\end{figure}

Nowadays, machine learning is one of the most popular research fields of modern science. The fundamental ingredient of machine learning systems is artificial neural networks (ANNs) since the most effective way of learning is done by ANNs. ANN is a computational model motivated by the biological nervous system. ANN is made up of computing units, called {\it neurons}. A schematic diagram of an ANN is given in Fig. \ref{fig:1}. 

\begin{figure}[H]
\centering
\includegraphics[width=3.4in]{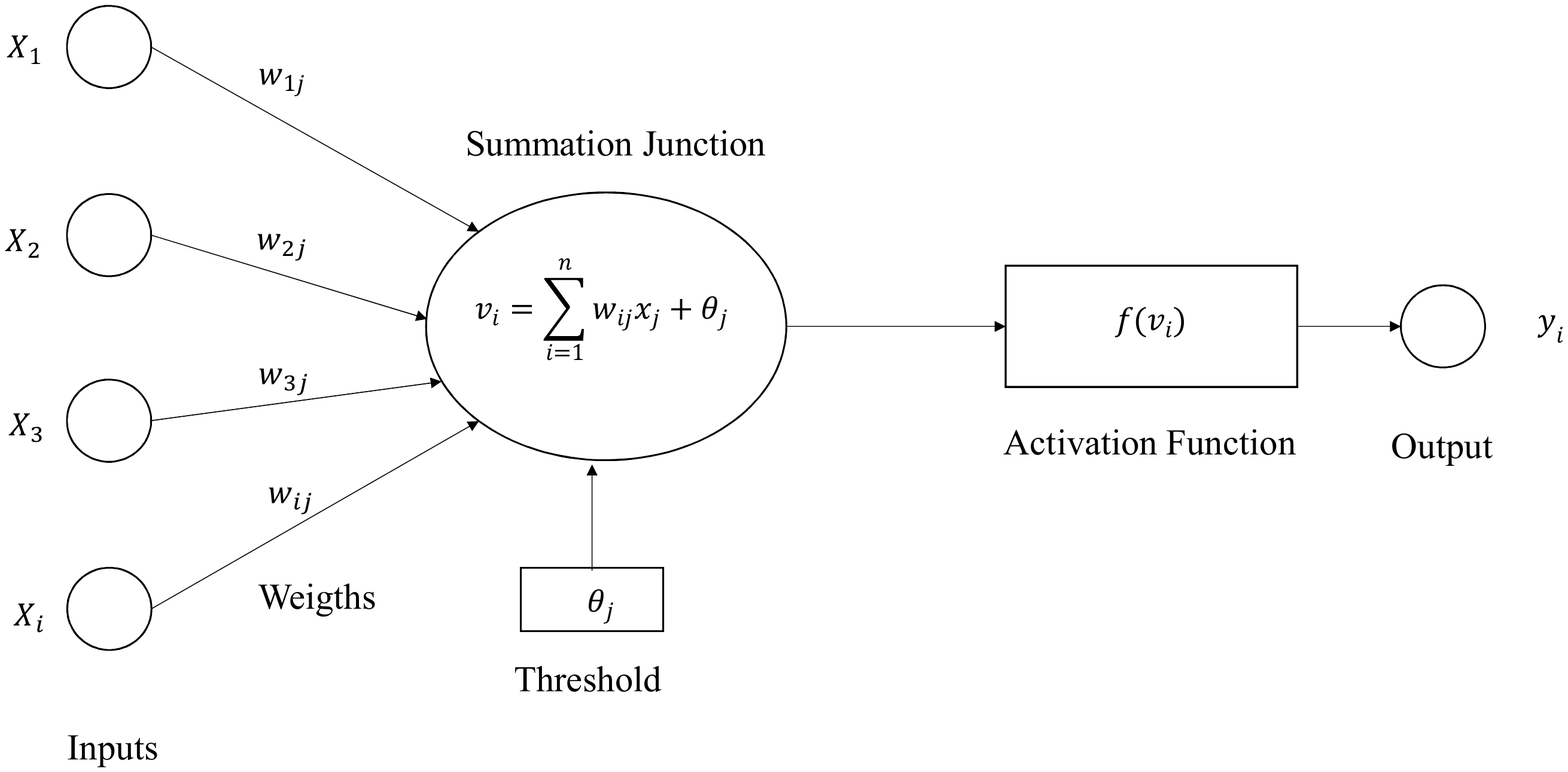}
\caption{\label{fig:2}  A model of single neuron}
\end{figure}

In this work, we use a multilayer perceptron (neuron) neural network (MLPN). A MLPN contains more than one layer of artificial neurons. These layers are connected to next layer but there is no connection among the neurons in the same layer. They are ideal tools for solving differential equations \cite{45}. A simple model of a neuron can be seen Fig. \ref{fig:2}.

Feed forward neural networks which are used in this present study, are the most used architectures because of their structural flexibility, good representational capabilities and a wide range of training algorithms available \cite{45}. All input signals are summed together as $z$ and the nonlinear activation function determines the output signal  $\sigma(z)$. We use a sigmoid function
\begin{equation}
\sigma(z)=\frac{1}{1+e^{-z}} \label{sigmoid}
\end{equation}
as an activation function since all derivatives of $\sigma(z)$ can be derived in terms of themselves. The information process can only be in one way in feed forward neural networks, from input layer(s) to output layer(s). The input-output properties of the neurons  can be written as 
\begin{eqnarray}
o_i&=&\sigma(n_i), \\
o_j&=&\sigma(n_j), \\
o_k&=&\sigma(n_k),
\end{eqnarray}
where $i$, $j$, and $k$ are for input, hidden, and output layers, respectively. Input to the perceptrons are given as
\begin{eqnarray}
n_i&=&(\text{Input signal to the neural network}), \\
n_j&=& \sum_{i=1}^{N_i} \omega_{ij}o_i+ \theta_j, \\
n_k&=& \sum_{i=1}^{N_j} \omega_{jk}o_j+ \theta_k,
\end{eqnarray}
where $N_i$ and $N_j$ represent the numbers of the units which belong to input and hidden layers respectively, $\omega_{ij}$ is the synaptic weight parameter connecting the neurons $i$ and $j$, and $\theta_j$ is threshold parameter for the neuron $j$ \cite{46}. The overall response of the network can be written as
\begin{equation}
o_k=\sum_{j=1}^{b_n} \omega_{jk}\sigma \left( \sum_{i=1}^{a_n}\omega_{ij}o_i +\theta_j \right)+ \theta_k. \label{eqn1}
\end{equation}
One can get the derivatives of $o_k$ with respect to the network parameters (weights and thresholds)  by differentiating Eqn. (\ref{eqn1}) as
\begin{eqnarray}
\frac{\partial o_k}{\partial \omega_{ij} }&=& \omega_{jk} \sigma^{(1)}(n_j)n_i,\\
\frac{\partial o_k}{\partial \omega_{jk} }&=& \sigma(n_j)\delta_{kk^\prime},\\
\frac{\partial o_k}{\partial \theta_j }&=& \omega_{jk} \sigma^{(1)}(n_j),\\
\frac{\partial o_k}{\partial \theta_{k^\prime} }&=&\delta_{kk^\prime}.
\end{eqnarray}

In order to obtain the spectra of pentaquark states, we consider of ANN application to a quantum mechanical system. We will follow the formalism which was formulated in \cite{40}. Consider the following differential equation
\begin{equation}
H\Psi(r)=f(r) \label{eqn2}
\end{equation}
where $H$ is a linear operator, $f(r)$ is a function and  $\Psi(r)=0$ at the boundaries. To solve this differential equation, it is possible to write a trial function as 
\begin{equation}
\Psi_t(\textbf{r})=A(\textbf{r})+B(\textbf{r}, \textbf{$\lambda$})N(\textbf{r}, \textbf{p}),
\end{equation}
which feeds a neural network with vector parameter  $\textbf{p}$ and $\textbf{$\lambda$}$ which are to be adjusted later. The parameter  $\textbf{p}$ stands for the weights and biases of the neural network.  $A(\textbf{r})$ and $B(\textbf{r}, \textbf{$\lambda$})$ should be conveniently specified in order to  $\Psi_t(\textbf{r})$ satisfies the boundary conditions regardless of the $\textbf{p}$ and $\textbf{$\lambda$}$ values.  In order to solve Eqn. (\ref{eqn2}), the collocation strategy can be utilized and it can be changed into a minimization problem as 
\begin{equation}
\underset{p,\lambda}{\min} \sum_i \left[ H\Psi_t(r_i)-f(r_i)  \right]^2. \label{dif2}
\end{equation}
 Eqn. (\ref{eqn2}) can be written as
\begin{equation}
H\Psi(r)=\epsilon \Psi(r) 
\end{equation}
with the boundary condition $\Psi(r)=0$. The trial solution can be written of the form
\begin{equation}
\Psi_t(r)=B(\textbf{r}, \textbf{$\lambda$})N(\textbf{r}, \textbf{p}),
\end{equation}
where $B(\textbf{r}, \textbf{$\lambda$}) =0$ at boundary conditions for a variety of $\lambda$ values. By discretizing the domain of the problem, Eqn. (\ref{dif2}) can be transformed into a minimization problem with respect to the parameters $\textbf{p}$ and $\textbf{$\lambda$}$
\begin{equation}
E(\textbf{p},\textbf{$\lambda$})=\frac{\sum_i \left[ H\Psi_t(r_i, \textbf{p},\textbf{$\lambda$})-\epsilon \Psi_t(r_i, \textbf{p},\textbf{$\lambda$})  \right]^2}{\int \vert \Psi_t \vert^2 d\textbf{r}},
\end{equation}
where $E$ is the error function and $\epsilon$ can be computed by
\begin{equation}
\epsilon=\frac{\int  \Psi_t^{\ast} H \Psi_t  d\textbf{r}}{\int \vert \Psi_t \vert^2 d\textbf{r}}.
\end{equation}

Consider a multilayer neural  network with $n$ input units, one hidden layer with $m$ units and one output.  For a given input vector
\begin{equation}
\textbf{r}=\left( r_1, \cdots,r_n  \right),
\end{equation}
the output of the network is 
\begin{equation}
N=\sum_{i=1}^m \nu_i \sigma(z_i),
\end{equation}
where 
\begin{equation}
z_i=\sum_{j=1}^n \omega_{ij}r_j+u_i.
\end{equation}
Here, $\omega_{ij}$ is the weight from input unit $j$ to hidden unit $i$, $\nu_i$ is the weight from hidden unit $i$ to output, $u_i$ is the bias of hidden unit $i$ and $\sigma(z)$ is the sigmoid function,  Eqn. (\ref{sigmoid}). The derivatives of output can be written as
\begin{equation}
\frac{\partial^k N}{\partial r^k_j}=\sum_{i=1}^m \nu_i \omega_{ij}^k \sigma_i^{(k)}
\end{equation}
where $\sigma_i=\sigma(z_i)$ and $\sigma^{(k)}$ is the $k$-th order derivative of the sigmoid. 

To obtain desired results, the first thing that ANN has to do is learning. The learning mechanism is the most important property of ANN. In this work, we used a feed forward neural network with a back propagation algorithm which is also known as delta learning rule. This learning rule is valid for continuous activation function, such as Eqn. \ref{sigmoid}. The algorithm is as follows \cite{47}:
\begin{enumerate}
\item[Step 1] Initialize the weights w from the input layer to the hidden layer
and weights v from the hidden layer to the output layer. Choose the
learning parameter (lies between 0 and 1) and error $E_{max}$. Initially error is taken as 0.
\item[Step 2] Train the network.
\item[Step 3] Compute the error value.
\item[Step 4] Compute the error signal terms of the output layer and the hidden layer.
\item[Step 5] Compute components of error gradient vectors.
\item[Step 6] Check the weights if they are properly modified.
\item[Step 7] If $E=E_{max}$ terminate the training session. If not, go to step 2
with $E \to 0$ and initiate a new training.
\end{enumerate}

\begin{figure}[H]
\centering
\includegraphics[width=3.4in]{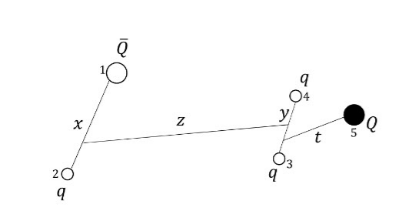}
\caption{\label{fig:3} Quark configuration with Jacobi coordinates \cite{25}}
\end{figure}

We parametrize trial function as
\begin{equation}
\phi_t(r)=r e^{-\beta r^2} N(r, \textbf{u}, \textbf{w}, \textbf{v}), ~ \beta >0 \label{wave}
\end{equation}
where $N$ denotes the feed forward artificial neural network with one hidden layer and $m$ sigmoid hidden units with
\begin{equation}
N(r, \textbf{u}, \textbf{w}, \textbf{v})=\sum_{j=1}^m \nu_j \sigma(\omega_j r+ u_j).
\end{equation}
The minimization problem becomes as
\begin{equation}
\frac{\sum_i \left[ H\phi_t(r_i)-\epsilon \phi_t(r_i)  \right]^2 }{\int \vert \phi_t(r) \vert ^2 dr}.
\end{equation}
We solved Schrödinger equation in the interval $0<r<1 ~ \text{fm}$ using 250 equidistant points with $m=10$. The wave function Eqn. (\ref{wave}) can accommodate the observed meson and baryon spectra. It is obvious that, the wave functions for mesons and baryons are different from the pentaquarks. In the case of pentaquark states, the wave function contain not only the spatial part but also spin, color and isospin parts.  In order to solve 5-body problem, Jacobi coordinates can be used \cite{25}:
\begin{eqnarray}
\vec{x}&=&\vec{r}_2-\vec{r}_1, ~\vec{y}=\vec{r}_4-\vec{r}_3, ~ \vec{t}=\vec{r}_5-\frac{\vec{r}_3+\vec{r}_4}{2}, \\ 
\vec{z}&=&\frac{\sum_{i=1}^2m_i\vec{r}_i}{\sum_{i=1}^2m_i}-\frac{\sum_{i=3}^5m_i\vec{r}_i}{\sum_{i=3}^5m_i}, ~ \vec{R}=\frac{\sum_{i=1}^5m_i\vec{r}_i}{\sum_{i=1}^5m_i}. 
\end{eqnarray}
Quark arrangements with this coordinates are shown in Fig. \ref{fig:3}. 

In this work we use the wave function of Ref. \cite{25} which reads as
\begin{eqnarray}
\Psi &=&\sum_{\alpha} \psi_{\alpha} \left(\vec{x},\vec{y},\vec{z},\vec{t} \right) \vert \alpha \rangle , \\ 
\psi_{\alpha} \left(\vec{x},\vec{y},\vec{z},\vec{t} \right)&=&\sum_i \gamma_{\alpha,i}  \exp \left(-\tilde{X}^\dagger \cdot A_{\alpha,i} \cdot X/2 \right), \label{pentawave}
\end{eqnarray}
where $\vert \alpha \rangle$ is color spin state, $A_{\alpha,i}$ are $4 \times 4$ positive definite matrices whose elements are the range parameters, and $\tilde{X}^\dagger=\left\lbrace \vec{x},\vec{y},\vec{z},\vec{t} \right\rbrace$. Color states are calculated using the SU(3) Clebsch-Gordan coefficients using the algorithm given in \cite{48}. Taking into account of spin, there are 5 independent spin arrangements for $S=1/2$ resulting 15 color-spin states $\vert \alpha \rangle$, 4 spin states for $S=3/2$ resulting 12  color-spin states, one spin state for $S=5/2$ resulting 3 color-spin sates. For isospin, there are two linearly independent isospin 1/2 vectors and one isospin 3/2 vector. For further discussion of color, spin and isospin, see Ref. \cite{26}. The range parameters of $A_{\alpha,i}$ in the wave function can be used to minimize the energy. For this purpose we parametrize Eqn. (\ref{pentawave}) as
\begin{equation}
\phi_t(x_i)= \sum_i \gamma_{\alpha,i}  \exp \left(-\tilde{X}^\dagger \cdot A_{\alpha,i} \cdot X/2 \right)   \vert \alpha \rangle N(x_i, \textbf{u}, \textbf{w}, \textbf{v}),
\end{equation}
and the minimization problem becomes as
\begin{equation}
\frac{\sum_i \left[ H\phi_t(x_i)-\epsilon \phi_t(x_i)  \right]^2 }{\int \vert \phi_t(x_i) \vert ^2 dx_i}.
\end{equation}

Before solving 5-body Schrödinger equation, some remarks are should be made. At first, the quark configuration in Fig. \ref{fig:3} represents asymptotic thresholds. In this configuration, pentaquark state is composed of an anticharmed meson and a charmed baryon. Asymptotic thresholds mean nominal reachable value as possible, summing the contribution of all quarks. They are reached when the range parameters of the trial function with the Jacobi coordinate of $\vec{z}$ vanish. 

The second point is that mass spectrum depend on the choice of the Hamiltoniand and trial function. In Ref. \cite{26}, the authors used a very similar Hamiltonian 
\begin{equation}
H=\sum_i \left( m_i+\frac{\bf{p}_i^2}{2m_i} \right)-\frac{3}{16}\sum_{i<j} \tilde{\lambda}_i \tilde{\lambda}_j V_{ij}(r_{ij})
\end{equation}
where $T_G$ is the kinetic energy of the center-of-mass system and $V_{ij}(r_{ij})$ potentials of \cite{44}, and with a different wave function. They calculated threshold energies with this Hamiltonian.
To test the choice of Hamiltonian, they used also AL1 potential of \cite{44} and found that the results of five-body calculations are essentially not modified. 	

Based on these arguments, we solved Schrödinger equation in the interval $0<x_i<1 ~ \text{fm}$ using 250 equidistant points with $m=10$.

\section{\label{sec:level3}Results and Discussion}
At first step, we calculated the masses of heavy mesons and baryons with all potentials with the wave function given in Eqn. (\ref{wave}). The results are given  in Table \ref{tab:table2}.
\begin{table}[H]
\caption{\label{tab:table2}Calculated masses of heavy mesons and baryons. All results are in MeV.}
\begin{ruledtabular}
\begin{tabular}{cccccccc}
Meson&Exp.&AL1&AP1& AL2 &AP2 \\
\hline
$\eta_c$& $2983$ & $2986$ & $2975$ &$2978$ &$2983$\\
 $J/\psi$& $3096$ & $3095$ & $3100$ &$3091$ &$3096$\\
$\bar{D}$& $1869$ & $1862$ & $1876$ &$1860$ &$1868$\\ 
$\bar{D}^\ast$& $2007$ & $2014$ & $2015$ &$2019$ &$2000$\\ 
\hline
Baryon&&&&  & \\
\hline
$N$& $938$ & $943$ & $932$ &$936$ &$946$\\ 
$\Lambda_c$& $2286$ & $2285$ & $2290$ &$2283$ &$2279$\\ 
$\Sigma_c$& $2455$ & $2471$ & $2463$ &$2475$ &$2482$\\
$\Sigma_c^\ast$& $2520$ & $2525$ & $2541$ &$2534$ &$2533$\\  
\end{tabular}
\end{ruledtabular}
\end{table}
One interesting point is that the potential (Eqn. (\ref{potential})) which have a simple form (has no many-body forces and tensor forces) reproduced masses of the observed states quite good. Motivated from these results, we obtained mass values of the newly observed pentaquark states according to their quantum numbers. Table \ref{tab:table3} shows the results of $J^P=1/2^-$ case and Table \ref{tab:table4} shows $J^P=3/2^-$ case, respectively.

\begin{table}[H]
\caption{\label{tab:table3}Calculated masses of pentaquark states for $J^P=1/2^-$. All results are in MeV. }
\begin{ruledtabular}
\begin{tabular}{cccccccc}
 State&Mass&AL1&AP1& AL2 &AP2 \\
\hline
$P_c(4312)$& $4311.9 \pm 0.7^{+6.8}_{-0.6}$ & $4314$ & $4317$ &$4320$ &$4312$\\
 $P_c(4440)$& $4440.3 \pm 1.3^{+4.1}_{-4.7}$ & $4360$ & $4371$ &$4372$ &$4374$\\
 $P_c(4457)$& $4457.3 \pm 0.6^{+4.1}_{-1.7}$ & $4390$ & $4388$ &$4395$ &$4392$\\ 
\end{tabular}
\end{ruledtabular}
\end{table}

\begin{table}[H]
\caption{\label{tab:table4}Calculated masses of pentaquark states for $J^P=3/2^-$. All results are in MeV.}
\begin{ruledtabular}
\begin{tabular}{cccccccc}
 State&Mass&AL1&AP1& AL2 &AP2 \\
\hline
$P_c(4312)$& $4311.9 \pm 0.7^{+6.8}_{-0.6}$ & $4371$ & $4382$ &$4377$ &$4369$\\
 $P_c(4440)$& $4440.3 \pm 1.3^{+4.1}_{-4.7}$ & $4441$ & $4445$ &$4439$ &$4445$\\
 $P_c(4457)$& $4457.3 \pm 0.6^{+4.1}_{-1.7}$ & $4456$ & $4458$ &$4450$ &$4457$\\ 
\end{tabular}
\end{ruledtabular}
\end{table}

It can bee seen from Tables \ref{tab:table3} and \ref{tab:table4}  that the mass of $P_c(4312)$ of four potentials with the quantum number assignment $J^P=\frac{1}{2}^-$ is more favourable than the quantum number $J^P=\frac{3}{2}^-$. On the other hand, the mass of  $P_c(4440)$ and $P_c(4457)$ with the quantum number assignment $J^P=\frac{3}{2}^-$ is more favourable than the the quantum number assignment $J^P=\frac{1}{2}^-$. All the potentials reproduced rather well the experimental data. 

In addition to the observed states, there will exist three six states with $J^P=1/2^-$ and $J^P=3/2^-$. We also calculated their mass values which are shown in Table \ref{tab:table5} for $J^P=1/2^-$ and in Table \ref{tab:table6} $J^P=3/2^-$, respectively. These states are denoted as $P_i$, where $i=1,\cdots 6$.

\begin{table}[H]
\caption{\label{tab:table5}Predicted masses of pentaquark states for $J^P=1/2^-$. All results are in MeV.}
\begin{ruledtabular}
\begin{tabular}{cccccccc}
 State&AL1&AP1& AL2 &AP2 \\
\hline
$P_1$& $3978$ & $3964$ &$4005$ &$3994$\\
 $P_2$& $4021$ & $4015$ &$4039$ &$4028$\\
 $P_3$&  $4075$ & $4059$ &$4051$ &$4062$\\ 
\end{tabular}
\end{ruledtabular}
\end{table}

\begin{table}[H]
\caption{\label{tab:table6}Predicted masses of pentaquark states for $J^P=3/2^-$. All results are in MeV.}
\begin{ruledtabular}
\begin{tabular}{cccccccc}
 State&AL1&AP1& AL2 &AP2 \\
\hline
$P_4$&  $4099$ & $4102$ &$4114$ &$4089$\\
 $P_5$&  $4125$ & $4120$ &$4130$ &$4118$\\
 $P_6$&   $4154$ & $4162$ &$4165$ &$4177$\\ 
\end{tabular}
\end{ruledtabular}
\end{table}

Two of these states lie below the $J/\psi p$ threshold and one of them is above for $J^P=1/2^-$ and three of them are slightly above the $J/\psi p$ threshold for $J^P=3/2^-$. This may require a different strategy for observing these states. A further detailed study of the $J/\psi p$ invariant mass spectrum can enlighten the status of these states.

The method of ANN for solving differential and eigenvalue equations include a trial function \cite{49}.  A trial function can be written as a feed forward neural network which includes adjustable parameters (weights and biases) and eigenvalue is refined to the existing solutions by training the neural network.  As mentioned in  Ref. \cite{26}, if a wave function results for a multiquark configuration an energy as $E=100 ~ \text{MeV}$ below the lowest threshold, it can represent the exact solution of the system. Besides this, an energy $E=100 ~ \text{MeV}$ above one of the threshold puts a question mark about the wave function and the model for describing the system. The relevant thresholds had been calculated in Ref. \cite{25} as $4329 ~ \text{MeV}$ for $D\Sigma_c$ with $I(J^P)=\frac{1}{2}(\frac{1}{2})^-$ and $4483 ~ \text{MeV}$ for $D^\ast \Sigma_c$ with $I(J^P)=\frac{1}{2}(\frac{3}{2})^-$. Our mass values are below at the order of $50 ~ \text{MeV}$ of the relevant thresholds which means trial function of this work represents the 5-body structure quite good.

The LHCb result could be an important sign to understand the heavy quark spin symmetry (HQSS). In the limit where the masses of heavy quarks are taken to infinity, the spin of the quark decouples from the dynamics which refers the strong interactions in the system are independent of the heavy quark spin. This implies that the states that differ only in the spin of the heavy quark, $i.e.$ states in which the rest of the system has the same total angular momentum, should be degenerate. This is also the case for single heavy baryons like $\Sigma_c^{\ast}$ $\Sigma_b^{\ast}$ and called heavy quark spin (HQS) multiplet structure. It is shown in Ref. \cite{38,39} that the HQS multiplet structure predicts a state near $\bar{D}^\ast \Sigma_c^\ast$ threshold with $J^P=5/2^-$. $\bar{D}^\ast \Sigma_c^\ast$ threshold with $J^P=5/2^-$ was calculated in Ref. \cite{25} as $4562 ~ \text{MeV}$. Our mass estimation for this state is shown in Table \ref{tab:table7}.

\begin{table}[H]
\caption{\label{tab:table7}Mass prediction of pentaquark state for $J^P=5/2^-$. All results are in MeV.}
\begin{ruledtabular}
\begin{tabular}{cccccc}
 &AL1&AP1& AL2 &AP2 \\
\hline
Mass & $4478$ & $4469$ &$4460$ &$4461$\\

\end{tabular}
\end{ruledtabular}
\end{table}
It should be also noted that a $5/2^-$ $\bar{D}^\ast \Sigma_c^\ast$ state does not couple to $J/\psi p$ in $S$- wave therefore it is not expected to give a peak in the LHCb \cite{39}. In fact, it is the phase space rather than partial wave dependence whether a state can produce a peak or not. The $J/\psi p$ threshold is around $4040~ \text{MeV}$ which is far below the mass of the $P_c$ state. Given a sufficiently large coupling, it can produce a peak in the $J/\psi p$ invariant mass spectrum even though the high partial wave is large. So there is still enough room to observe this state.

\section{\label{sec:level4}Summary and Concluding Remarks}
Inspired by the recent observation of the hidden-charm pentaquark states, we solved 5-body Schrödinger equation in the nonrelativistic quark model framework. We used a nonrelativistic quark model using the potentials proposed in \cite{44}. These potentials reproduced the experimental ground state masses of some mesons and baryons as a demonstration of the method. We used ANN method to get the solution of the 5-body Schrödinger equation. 

We gave a prediction of quantum numbers for these newly observed pentaquarks. The quantum number assignments for $P_c(4312)$, $P_c(4440)$, and $P_c(4457)$ of this work are in agreement with \citep{33,34,36,38}. Since the spin and parity numbers are not determined in the LHCb report, the other $J^P$ assignments can not be excluded. For example the $P_c(4440)$ and $P_c(4457)$ states can be explained as $5/2^+$ and $5/2^-$ $\bar{D}^\ast \Sigma_c$ state \cite{50}. Partial wave analysis in the experimental data is critical to enlighten the internal structures of these exotic states. 

We also calculated the mass for $5/2^-$ $\bar{D}^\ast \Sigma_c^\ast$ state which is a prediction of heavy quark spin multiplet structure. The average mass value of four estimations is roughly $95 ~ \text{MeV}$ below the relevant threshold. Searching this missing HQS partner or partners is an important task for future experiments.

Within framework of Hamiltonian in this work, a molecular picture for the newly observed pentaquark states can not be concluded or excluded. Both the mass uncertainties and decay properties should be studied. The kinematic vicinity of the observed pentaquark states to the charmed meson-charmed baryon thresholds does not corroborate that they are molecules. In Ref. \cite{51}, it is found that masses and decay properties of the $P_c(4457)^+$, $P_c(4440)^+$, and $P_c(4312)^+$ can be understood if one treats them as $J^P=3/2^-$, $J^P=1/2^-$ and $J^P=3/2^-$, compact pentaquark states, respectively. These properties can also be obtained in the molecule picture assuming them as $J^P=3/2^- ~(1/2^-)$, $J^P=1/2^- ~(3/2^-)$, and $J^P=(1/2^-)$ S-wave states, respectively. 

\begin{acknowledgments}
The author thanks to C. Hanhart and anonymous referee for their valuable comments in the revised version of this paper.
\end{acknowledgments}


\end{document}